\documentclass[a4paper,11pt]{article}
\pdfoutput=1
\usepackage[normalem]{ulem}
\usepackage[utf8]{inputenc}
\usepackage{cite}
\usepackage[left=2cm,right=2cm,top=2cm,bottom=3cm]{geometry}
\usepackage{xcolor}
\usepackage{float}

\usepackage{epsf}
\usepackage{graphicx}

\usepackage{amsmath}
\usepackage{amssymb}
\usepackage{mathrsfs}
\usepackage{slashed}
\usepackage{mathtools}

\usepackage{amsfonts}
\usepackage{xfrac} %for nice small fractions: sfrac
\usepackage{booktabs} % for nice tables
\usepackage{soul} % for striking out \st{text}

\usepackage{color}
\definecolor{cred}{RGB}{180,50,40}
\definecolor{purple}{RGB}{180,90,180}
\definecolor{darkgreen}{RGB}{0, 100, 0}

\usepackage[
colorlinks=true
,urlcolor=black
,anchorcolor=black
,citecolor=black
,filecolor=black
,linkcolor=black
,menucolor=black
,linktocpage=true
,pdfproducer=medialab
,pdfa=true
]{hyperref}

\def\bsll{b \to s\ell\ell}
\def\bclnu{b \to c\ell\nu}
\def\rkrks{R_{K^{(\ast)}}}
\def\rdrds{R_{D^{(\ast)}}}
\def\rk{R_{K}}
\def\rks{R_{K^\ast}}

\begin{document}
\vspace*{1cm}
\allowdisplaybreaks
\setcounter{footnote}{0}
\begin{center}
\vspace*{1mm}
%%%%%%%%%%%%%%%%%%%%%
\vspace{1cm}
{\Large\bf
Leptoquarks facing flavour tests and $\pmb{b\to s\ell\ell}$ after Moriond 2021}

\vspace*{0.8cm}

{\bf J.~Kriewald $^{a}$, C.~Hati $^{b}$, J. Orloff $^{a}$ and A.~M.~Teixeira $^{a}$}

\vspace*{.5cm}
$^{a}$Laboratoire de Physique de Clermont (UMR 6533), CNRS/IN2P3,\\
Univ. Clermont Auvergne, 4 Av. Blaise Pascal, F-63178 Aubi\`ere Cedex,
France

\vspace*{.2cm}
$^{b}$ Physik Department T70, Technische Universit\"at M\"unchen,\\
James-Franck-Stra{\ss}e 1, D-85748 Garching, Germany

\end{center}

\vspace*{2mm}
\begin{abstract}
\noindent
In view of the emerging hints for the violation of lepton flavour universality in several $B$-meson decays, we conduct a model-independent study (effective field theory approach) of several well-motivated new physics scenarios. 
Taking into account the most recent LHCb data, we provide updates to Wilson coefficient fits for numerous popular new physics hypotheses.
We also consider a promising model of vector leptoquarks,
which in addition to explaining the $B$-meson decay anomalies ($R_{K^{(*)}}$ and $R_{D^{(*)}}$) would have an extensive impact for numerous flavour observables. We identify 
promising decay modes allowing to (indirectly) probe such an extension: these include
positive signals (at Belle II or LHCb) for $\tau\to \phi \mu$, $B_{(s)}$-meson decays to $\tau^+ \tau^-$ and $\tau^+ \mu^-$ ($\tau^+ e^-$) final states, as well as an observation of certain charged lepton flavour violation transitions at COMET and Mu2e. We also argue how the evolution of the experimental determination of $R_{D^{(*)}}$ can prove instrumental in falsifying a vector
leptoquark explanation of the anomalous $B$-meson decay data.
\end{abstract}

\section{Introduction}
In the Standard Model (SM), charged leptons are only distinguishable due to their masses. In particular, all  
electroweak couplings to gauge bosons are blind to lepton flavour, leading to an accidental symmetry called lepton flavour universality (LFU), whose validity has been determined to a very high accuracy for instance in $Z\to \ell^+\ell^-$ and $W^\pm\to \ell^\pm \nu$ ($\ell = e, \mu, \tau$) decays~\cite{PDG}.

However, during the last decade, hints on the violation of LFU in $\bclnu$ and $\bsll$ decays have begun to emerge, in mounting tension with respect to the SM expectations.
In particular, measurements of the ``theoretically clean'' ratios of branching ratios ${R_{D^{(\ast)}}} = \mathrm{BR}(B\to D^{(\ast)}\tau\nu)/\mathrm{BR}(B\to D^{(\ast)}\ell\nu)$~\cite{Belle:2019rba} and ${R_{K^{(\ast)}}} = \mathrm{BR}(B\to K^{(\ast)}\mu\mu)/\mathrm{BR}(B\to K^{(\ast)}ee)$~\cite{rkrks} deviate around $2 - 3\,\sigma$ from their theoretical predictions, which are, up to phase space suppression, expected to be unity in the SM.
Current averages of experimental measurements and the SM predictions can be found in Table~\ref{tab:data}.
\begin{table}[h!]

\begin{center}
{\small
\begin{tabular}{|c|c|c|c|c|}
\hline
     &  $R_{K}$ & $R_{K^{*}}$ & $R_{D}$ & $R_{D^{*}}$\\
     \hline
SM prediction  &  $\simeq 1$ & $\simeq 1$ 
& $0.299 \pm 0.003$ & $0.258 \pm 0.005$ \\
\hline
Experiment     & $0.845 \pm 0.06$ & $0.69 \pm 0.12$  
& $0.340 \pm 0.030$ & $0.295 \pm 0.014$\\
\hline
\end{tabular}
}
\label{tab:data}
\caption{SM predictions and (averages of) experimental measurements of the ``theoretically clean'' LFU observables.}
\end{center}

\end{table}

Additionally, measurements of angular observables in $B^{0,+}\to K^\ast \mu^+\mu^-$ decays exhibit (local) deviations of $2-3\,\sigma$ in several $q^2$ bins.
These measurements~\cite{angular} have been recently updated, confirming and strengthening the currently preferred NP hypotheses.

Very recently, the LHCb Collaboration updated~\footnote{Due to an improved determination of the $f_s/f_d$ fragmentation in $B_{(s)}$-decays LHCb also updated their measurement on $B_{(s)}\to \mu^+\mu^-$ decays. However, we have decided to refrain from ``digitising''the plots presented on slides, and thus cannot take into account the update in a consistent manner for the present fits.} their measurement of $\rk = 0.846^{+0.044}_{-0.041}$~\cite{Aaij:2021vac} with the deviation to the SM prediction now reaching $3.1\,\sigma$, thus providing the first {\it evidence} of LFU violation~\footnote{For an (animated) overview of how the fit to $\bsll$ data has evolved upon inclusion of the new measurements see \href{http://moriond.in2p3.fr/2021/EW/slides/ani_fit_evo.mp4}{\textcolor{blue}{http://moriond.in2p3.fr/2021/EW/slides/ani\_fit\_evo.mp4}}.}.

These curious, persisting tensions with the SM seem to indirectly hint towards the presence of new physics (NP), likely at the TeV-scale. 
Many different approaches have been explored to identify 
which (minimal) NP models better succeed in reconciling theoretical predictions with the experimental data. 
Before addressing the prospects of vector leptoquark (LQ) extensions of the SM (one of the most promising well motivated scenarios to simultaneously explain both anomalies), we will consider a model-independent effective field theory (EFT) based approach. This will allow to generically identify which classes of NP models offer the most appropriate content and interactions to explain the anomalous data. 

\section{EFT and global fits}
The EFT approach relies in parametrising NP effects in terms of higher-order non-renormalisable operators (vestigial traces in the low-energy theory of heavier  states, which are integrated out). Starting from the relevant subsets of the effective Lagrangian, cast in terms of the relevant semileptonic Wilson coefficients (WC) $C^{q q^\prime; \ell \ell^\prime }$ and effective operators, we
comment on how well-motivated scenarios for (sets of) $C^{q q^\prime; \ell \ell^\prime }$ become significantly favoured by current data.

\begin{figure}
    \centering
    \includegraphics[width=0.48\textwidth]{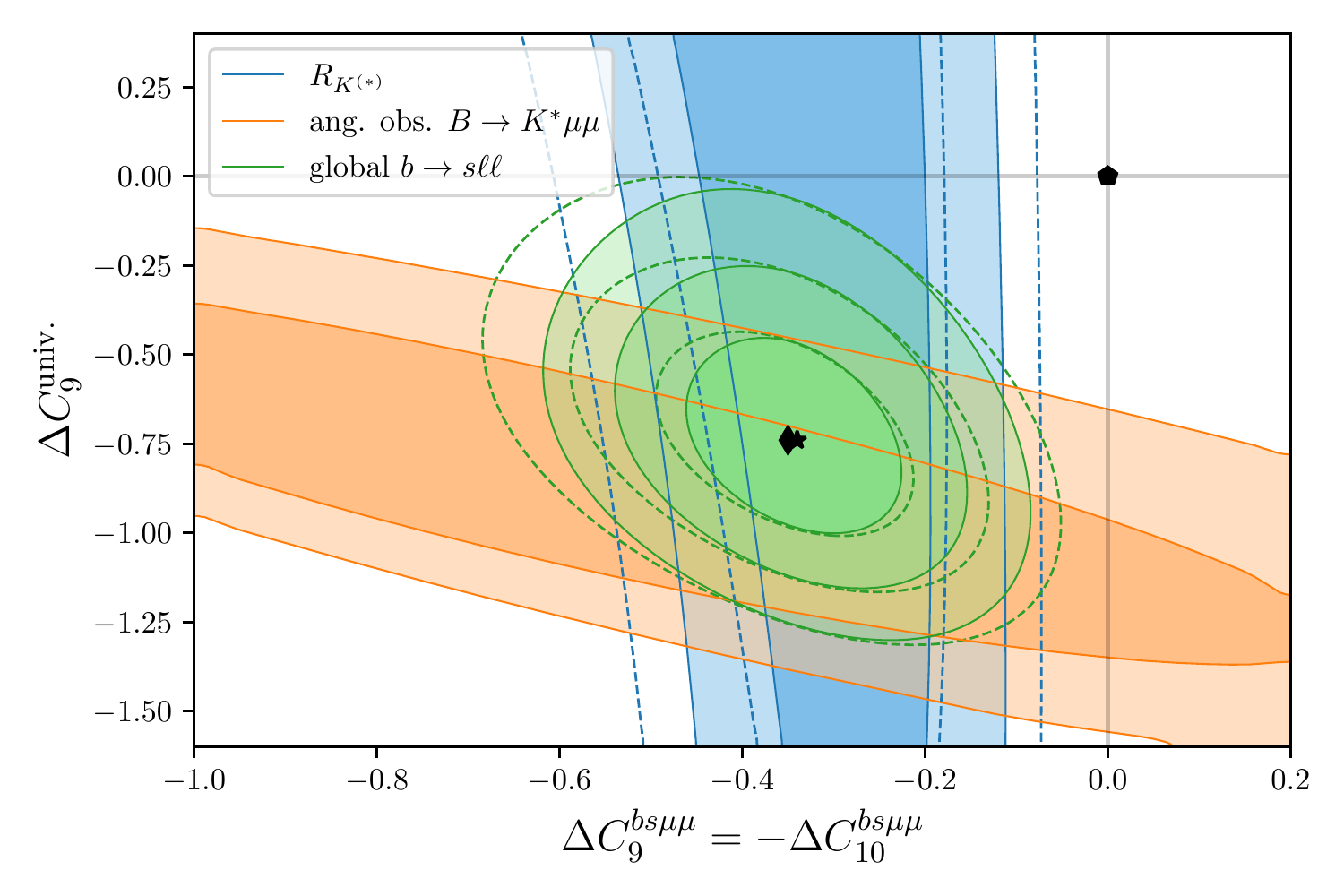}\includegraphics[width=0.48\textwidth]{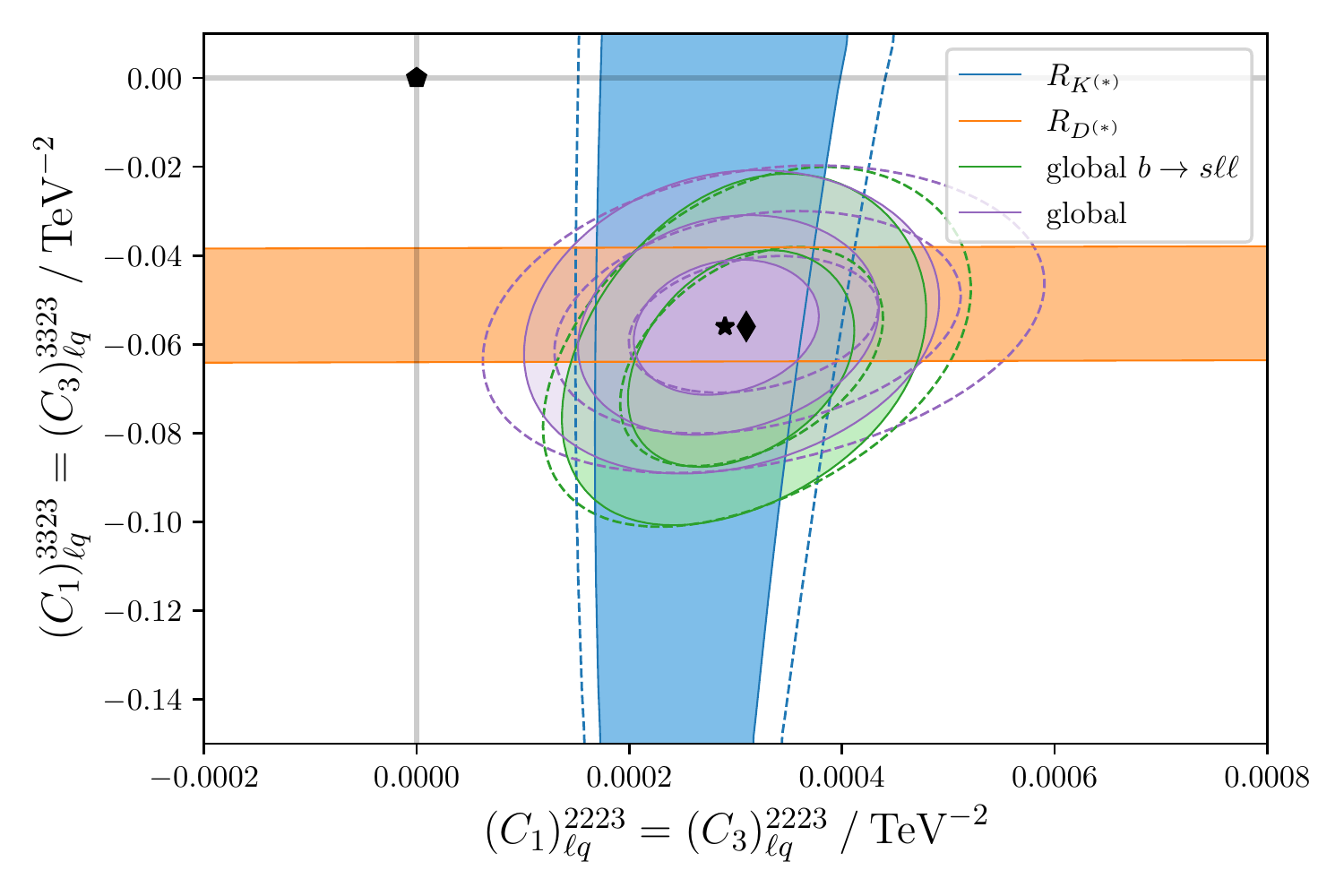}
    \caption{Likelihood contours (1$\sigma$, 2$\sigma$ (and 3 $\sigma$ for the global fit)) in the plane of WET and SM-EFT Wilson coefficients. Dashed contour lines denote the situation prior to the updated 2021 $\rk$ measurement, a pentagon the SM prediction, a diamond the ``old'' best fit point and a star the new best fit point upon inclusion of the updated $\rk$ data. {\bf Left:} Fit of WET coefficients at $4.8\,\mathrm{GeV}$ with the best fit point $\Delta C_9^{bs\mu\mu} = -0.34^{+0.08}_{-0.08}$ and $\Delta C_9^\mathrm{univ.} = -0.74^{+0.19}_{-0.17}$. {\bf Right:} Fit of SM-EFT coefficients at $2\,\mathrm{TeV}$ with the best fit point of $(C_1)_{\ell q}^{2223} = (2.9^{+0.6}_{-0.6})\times 10^{-4}\,\mathrm{TeV}^{-2}$ and $(C_1)_{\ell q}^{3323} = 0.056^{+0.01}_{-0.01}\,\mathrm{TeV}^{-2}$, with a pull of $7.9\sigma$ with respect to the SM prediction. Figures taken from Ref.~\cite{talk}.}
    \label{fig:fits}
\end{figure}

The subset of the effective Lagrangian for charged current $d_k\to u_j\ell\nu_i$ transitions is given by
\begin{equation}
    \mathcal L_{\mathrm{eff}} \simeq -\frac{4 G_F}{\sqrt{2}}V_{jk}\left[(1 + C_{V_L}^{jk\ell i})(\bar u_j \gamma_\mu d_k)(\bar \ell \gamma^\mu P_L \nu^i) \right] + \mathrm{H.c.}\,,
\end{equation}
where $V_{jk}$ are elements of the Cabibbo-Kobayashi-Maskawa (CKM) quark mixing matrix.
While the charged current anomalies $\rdrds$ can be explained with NP contributions to the left-handed vector coefficient $\mathcal C_{V_L}^{c b \tau \nu}$, the neutral current ones -- especially due to the deviations in the angular observables -- call upon a dedicated EFT analysis to identify which operator structure (or combination of structures) is 
preferred by experimental data.
A subset of the low energy effective Lagrangian in $\bsll$ transitions can be cast as
\begin{equation}
    \mathcal L_\mathrm{eff} \simeq \frac{4  G_F}{\sqrt{2}}V_{t d_j}V_{t d_i}^\ast\left[\frac{\alpha_e}{4\pi} C_9^{ij\ell\ell'}(\bar d_i\gamma^\mu P_L d_j)(\bar \ell \gamma_\mu \ell') + \frac{\alpha_e}{4\pi} C_{10}^{ij\ell\ell'}(\bar d_i\gamma^\mu P_L d_j)(\bar \ell \gamma_\mu\gamma_5 \ell')\right]\,.
\end{equation}
As it turns out, a very interesting NP scenario is given by considering LFU violating $V-A$ new physics contributions in $\Delta C_9^{bs\mu\mu} = - \Delta C_{10}^{bs\mu\mu}$, in addition to a LFU vector NP contribution denoted by $\Delta C_9^\mathrm{univ.}$ which is flavour universal and added to $C_9^{bs\mu\mu}$ and $C_9^{bsee}$. 
%universally. 
A fit of all available $\bsll$ data~\footnote{For a complete list of observables taken into account, see Appendix~B of Ref.~\cite{LQ2020}.}, including the updated measurement in $\rk$, leads to a $\sim 6.6 \sigma$ improvement with respect to the SM prediction. The best fit point is found to be given by $\Delta C_9^{bs\mu\mu} = -0.34^{+0.08}_{-0.08}$ and $\Delta C_9^\mathrm{univ.} = -0.74^{+0.19}_{-0.17}$, and shows a $\sim 3\sigma$ preference for non-vanishing NP contributions to the WCs~\cite{talk}, as can be seen by the likelihood contours 
in the plane of weak effective theories (WET) and SM-EFT Wilson coefficients, shown in the left panel of Figure~\ref{fig:fits}.

Interestingly, and as it has been pointed out in Ref.~\cite{Crivellin:2018yvo}, a universal contribution to $C_9^{bs\ell\ell}$ can be generated from renormalisation group (RG) running from sizeable semi-tauonic operators. 
In the right panel of 
Figure~\ref{fig:fits} we display the
likelihood~\footnote{Other than the $\bsll$ and $\rdrds$ data, the global likelihood also contains the binned branching fractions in exclusive $B\to D^{(\ast)}\ell\nu$ decays.}
contours in the plane of the semi-muonic and semi-tauonic $SU(2)_L$ singlet and triplet SM-EFT Wilson coefficients in $\bsll$. As can be seen, the $\bsll$ contours are not independent of the semi-tauonic operators, an effect of the RGE-induced universal contributions to $C_9^{bs\ell\ell}$ at the $b$-quark mass scale.

This correlation between the charged and neutral current anomalies at the high-energy scale strongly hints towards a combined explanation of both tensions in a minimal model.
For completeness, in Appendix~\ref{app:fits}
%of the arXiv version of this proceeding
we include updates on additional NP hypotheses.

Interestingly, the preferred structure of EFT operators (both in SM-EFT and WET) is naturally generated by the $SU(2)_L$-singlet vector LQ $V_1$, contributing to both charged and neutral current anomalous transitions at the tree-level, while evading stringent constraints from $d_i\to d_j \nu\bar\nu$ decays.

\section{Vector leptoquarks}
Gauge vector LQs, such as $V_1$, naturally arise in (grand) unified theories, specifically from quark-lepton unification, as it occurs in Pati-Salam models.
In our study of Ref.~\cite{Hati:2019ufv}, and  
instead of exploring a specific UV completion for $V_1$ leptoquarks, we chose to find requirements on the couplings of $V_1$ to the SM fermions in an effective way.
The subset of the relevant 
(left-handed~\footnote{Due to the absence of strong hints 
suggesting non-vanishing right-handed contributions to the WET operators, we restrict the LQ couplings to be left-handed.}) LQ couplings to SM fermions 
can be parametrised in a general way as
\begin{equation}
    \mathcal L \simeq \sum_{i,j,k,l = 1}^{3} V_1^\mu \left(\bar d_L^i\gamma_\mu K_L^{ik}\ell_L^k + \bar u_L^j V_{ji}^\dagger \gamma_\mu K_L^{ik} U_{kl}^P\nu_L^l \right) + \mathrm{H.c.}\,,
\end{equation}
where $K_L^{ij}$ denote the effective LQ couplings and $U^P$ is the Pontecorvo-Maki-Nakagawa-Sakata (PMNS) leptonic mixing matrix.
The matching of the LQ couplings (at the LQ mass scale, $m_{V_1} \simeq 1.5\,\mathrm{TeV}$) to the WCs singled out by the EFT analysis can be done as follows:
\begin{equation}
C^{ij;\ell \ell^{\prime}}_{9,10} = \mp\frac{\pi}{\sqrt{2}G_F\,\alpha_\mathrm{em}\,V_{3j}\,V_{3i}^{\ast} \,m_{V_1}^2}\left(K_L^{i
  \ell^\prime} \,K_L^{j\ell\ast} \right)\,,\ \ 
  C_{jk,\ell i}^{V_L} = \frac{\sqrt{2}}{4\,G_{F}\,m_{V_1}^2}\,
  \frac{1}{V_{jk}}\,
  (V\,K_{L}\, U^P)_{ji}\, K_{L}^{k\ell\ast}\,.
  \label{eqn:CV}
\end{equation}

\begin{figure}
    \centering
    \mbox{\raisebox{-2mm}{\includegraphics[width=0.48\textwidth]{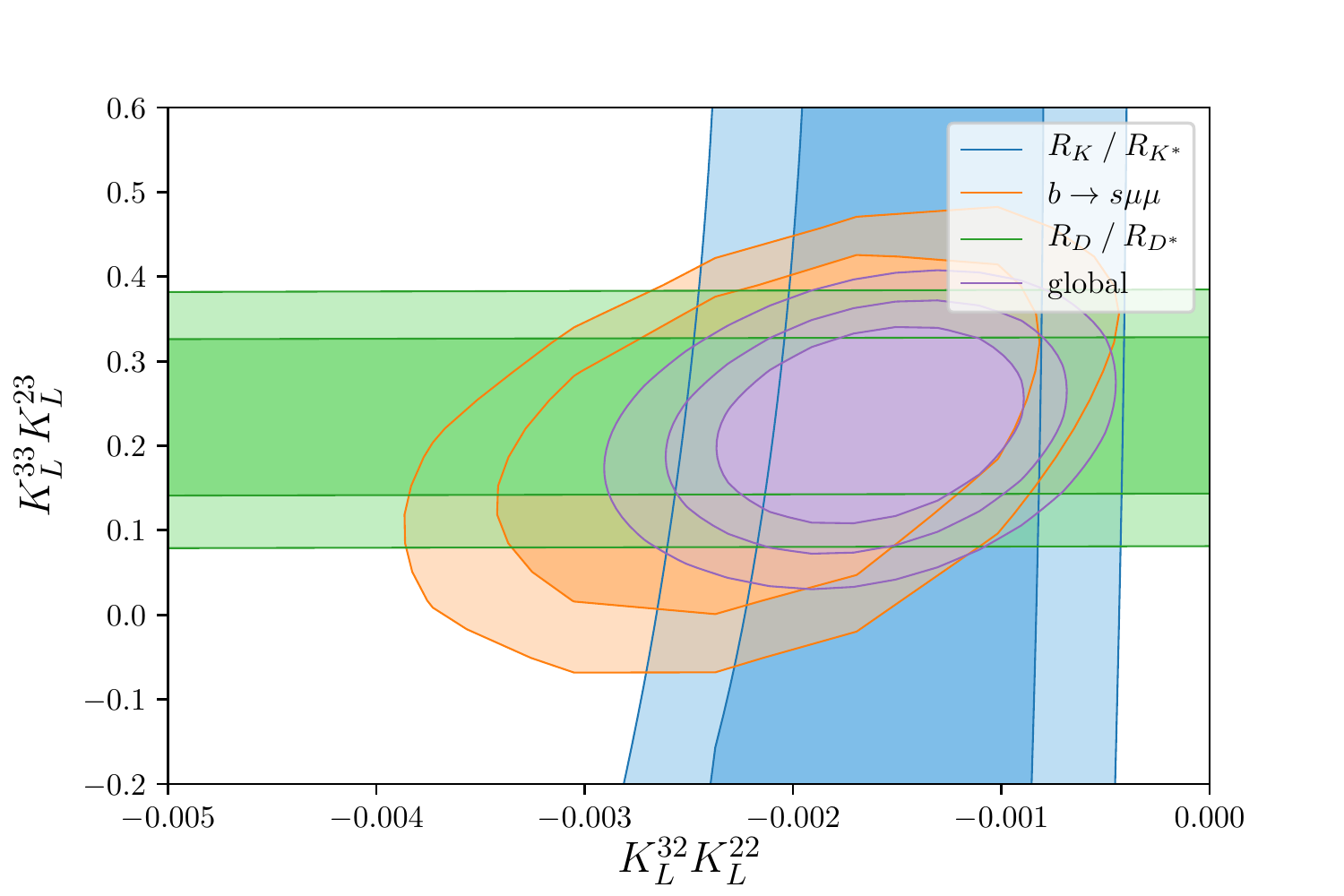}}
    \raisebox{-8mm}{\includegraphics[width=0.48\textwidth]{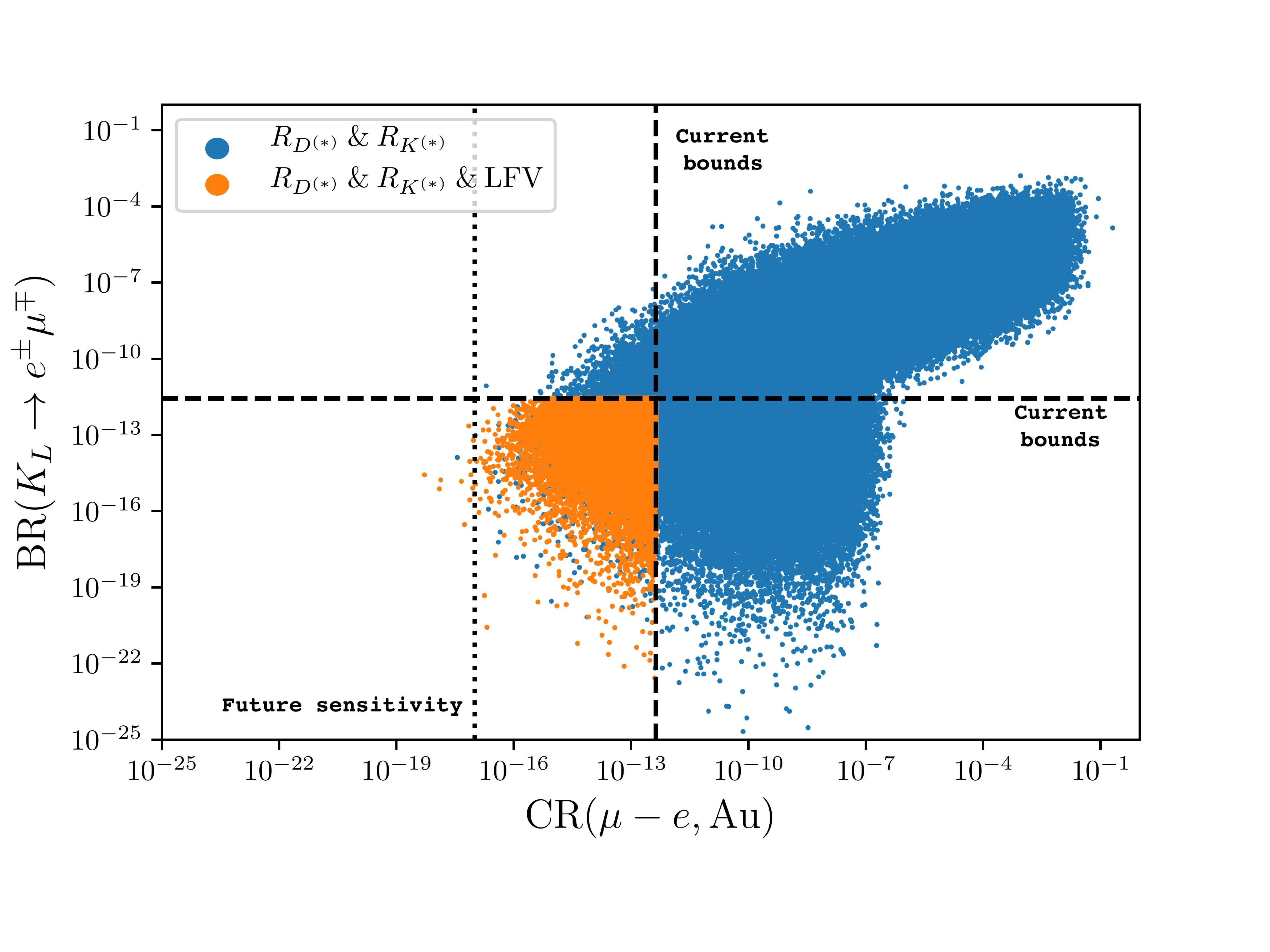}}}
    \caption{{\bf Left:} Fit of combinations of the dominant $V_1$ LQ couplings ($K_L^{3i}K_L^{2i}$) to the anomalous data. The likelihood contours correspond to 1 $\sigma$, 2 $\sigma$ and 3 $\sigma$. {\bf Right:} Generated Monte-Carlo samples accommodating the $B$ anomalies, displayed in the plane of the two most constraining observables, CR$(\mu-e, \mathrm{Au})$ and $K_L \to e^\pm \mu^\mp$. Blue points violate at least one LFV bound while orange points comply with {\it all} imposed constraints. Figures taken from Ref.~\cite{Hati:2019ufv}.}
    \label{fig:LQfit}
\end{figure}

In the left panel of Figure~\ref{fig:LQfit} we show likelihood contours in the plane of the dominant LQ couplings. Notice that one finds the same correlation between semi-muonic and semi-tauonic couplings in the likelihood contours preferred by $\bsll$ data as previously encountered in the EFT fits (cf. Figure~\ref{fig:fits}).

Sizeable $V_1$ couplings of second- and third-generation quarks to different lepton flavours can lead to strongly enhanced rates in lepton flavour violating (LFV) processes, such as $B\to K\tau\mu$, which put stringent constraints on the model's parameter space.
As we have shown in our analysis in~\cite{Hati:2019ufv}, in order to evade those bounds, effectively non-unitary couplings to the SM fermions are necessary. For instance, this can be achieved by introducing vector-like fermions, more specifically vector-like leptons, that mix with the SM fermions.
In our approach we have thus parametrised the LQ couplings via 12 unitary rotations incorporating the mixing between the SM leptons and 3 additional generations of vector-like $SU(2)_L$ doublet leptons~\footnote{Other representations of vector-like leptons are 
ruled out since the required mixing pattern would drastically impact $Z\ell\ell$ couplings already at the tree-level, thus violating experimental bounds on these precisely measured quantities.}.
In the right panel of Figure~\ref{fig:LQfit} we show results of a random scan where we vary the 12 mixing angles in the interval $[-\pi,\pi]$, presenting our results in the plane of the two most constraining LFV observables, namely $K_L\to \mu^\pm e^\mp$ decays and neutrinoless $\mu-e$ conversion in nuclei. The points displayed (blue) are in agreement with the anomalous data at the $3\,\sigma$ level; however, most of the latter points leads to the violation of at least one experimental LFV bound. Points complying with all imposed constraints (explaining $R_{K,D}^{(*)}$ and respecting all experimental bounds) are marked in yellow. Notice that most of the currently preferred parameter space can be probed by the upcoming experiments COMET and Mu2e~\cite{mutoe}, both dedicated to searching for neutrinoless $\mu-e$ conversion in Aluminium.

In a second - updated - phenomenological analysis we have now taken the 9 left-handed LQ couplings as independent parameters and we have fit them to more than 350 observables~\footnote{These include $\bsll$ and $\bclnu$ data, a large number of lepton flavour violating and conserving (semi-)leptonic ($b$, $c$ and $s$) meson and $\tau$ decays and several purely leptonic LFV processes. A complete list can be found in Ref.~\cite{LQ2020}.}, for three mass benchmark points of $m_{V_1}\in[1.5, 2.5, 3.5]\,\mathrm{TeV}$; this allows finding a region in the $9$-dimensional parameter space in which the $B$-anomalies can be explained while evading constraints from LFV processes~\cite{LQ2020}.
We assumed the posterior distribution of the couplings to be approximately gaussian and sampled them according their distribution. From the Monte-Carlo samples we have then computed posterior ranges for the observables around the best fit point(s), based on the currently by experimental data preferred parameter space.

\begin{figure}
    \centering
    \mbox{\hspace{-10mm}\raisebox{0mm}{\includegraphics[width=0.65\textwidth]{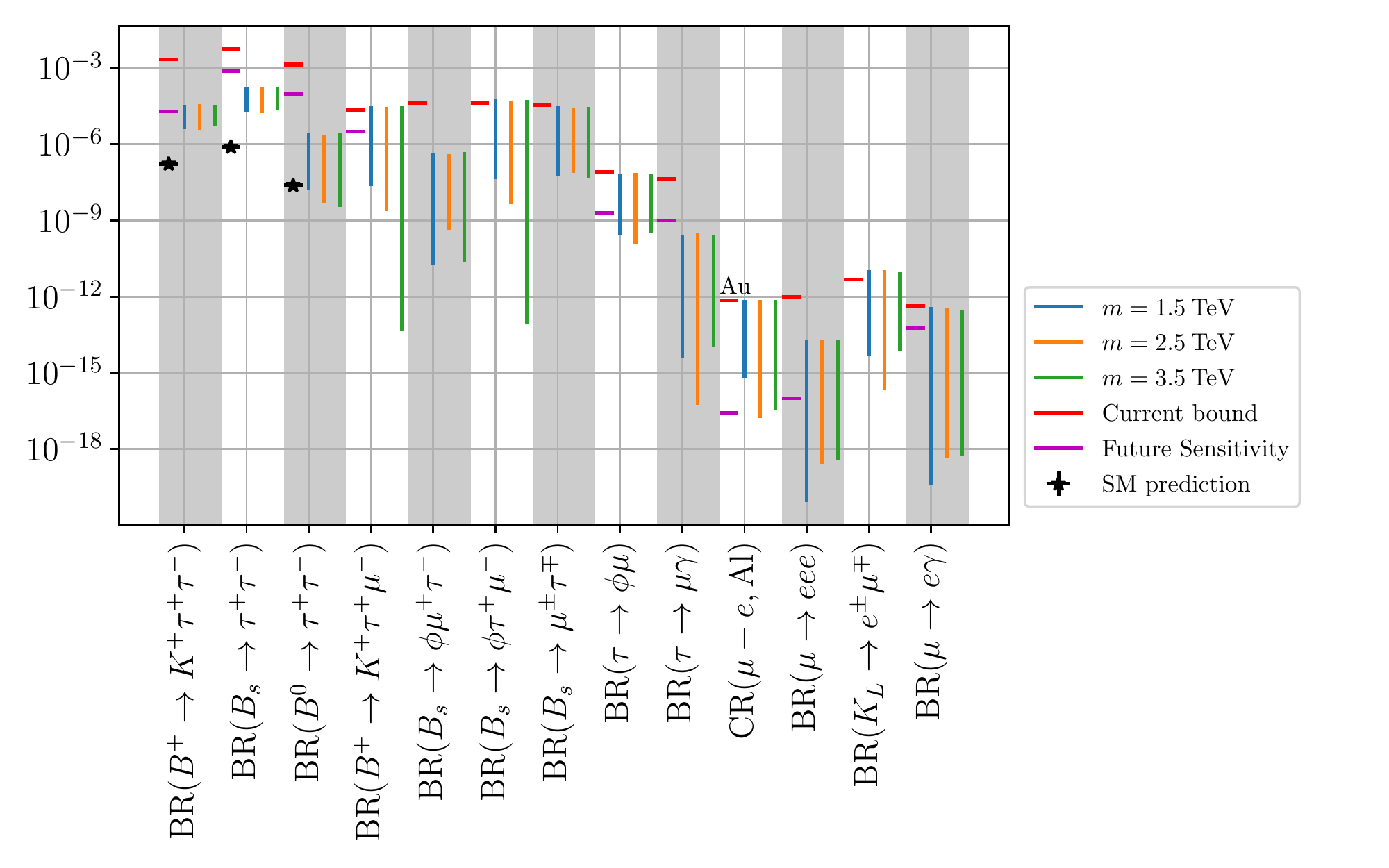}}
    {\hspace{-4mm}\raisebox{6mm}{\includegraphics[width=0.55\textwidth]{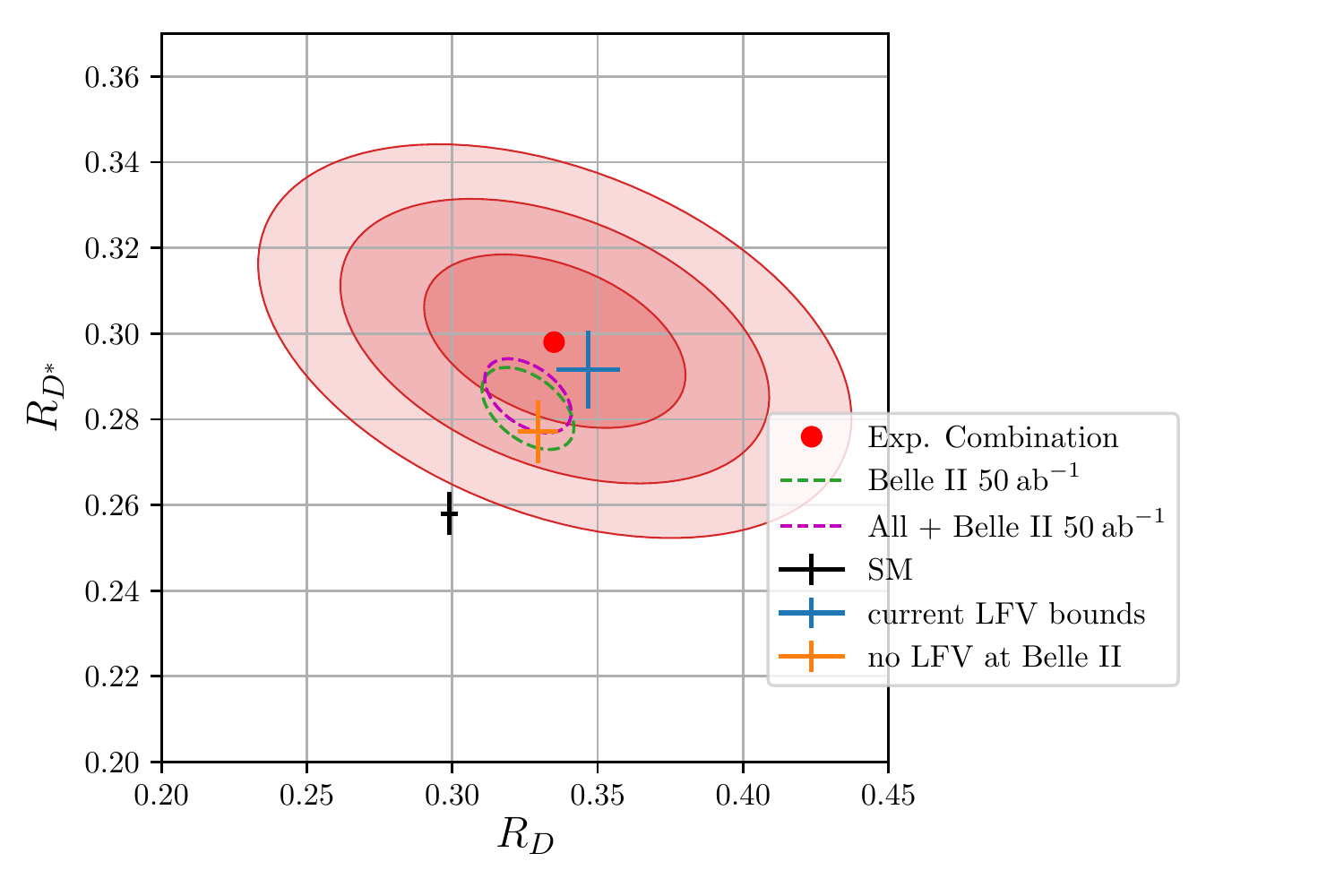}}}}
    \caption{{\bf Left:} Posterior ranges of the vector LQ predictions for several observables searched for by Belle II, COMET and Mu2e. {\bf Right:} Posterior ranges for LQ predictions for $\rdrds$ based on current (future) bounds on LFV processes relevant for the test of the vector LQ hypothesis presented in blue (orange). The predictions for different masses coincide. Also shown are the extrapolations of the current Belle measurement to the sensitivity of Belle II (dashed contourlines). Figures taken from Ref.~\cite{LQ2020}.}
    \label{fig:predictions}
\end{figure}

Several rare $B$-decays involving tau-leptons and nnumerous LFV tau-decays are expected to be searched for by the Belle II experiment~\cite{Kou:2018nap} with improved sensitivities.
Due to sizeable couplings of the LQ to $b, s$- and $c$-quarks, and to charged leptons, a priori we expect sizeable enhancements of $b\to s \tau^+\tau^-$ and LFV processes.
In the left panel of Figure~\ref{fig:predictions} we show the $1\sigma$ ranges of the posterior distributions for several such observables, together with current experimental limits and future sensitivities.
In particular, $B\to K\tau\mu$ and $\tau\to\phi\mu$ decays are promising channels at Belle II. Additionally, the predicted range of neutrinoless $\mu-e$ in Aluminium will be (almost) fully probed by COMET and Mu2e~\cite{mutoe}.

The results of this analysis are by no means a guarantee to discover LFV signals in those channels. 
Thus, we have furthermore studied the impact of {\it null} results in LFV channels at Belle II, Mu2e and COMET - thus replacing current LFV bounds with future sensitivities in our fit.
Results of such a hypothetical future fit are presented in the right panel of Figure~\ref{fig:predictions}, where we show predictions of the LQ model for $\rdrds$ based on current and future LFV bounds (blue and orange respectively).
Moreover, we have extrapolated the current measurement of $\rdrds$ by Belle to the anticipated accuracy of Belle II with $50\,\mathrm{ab}^{-1}$ integrated luminosity, and combined it with all other measurements available, denoted via the green and purple (dashed) ellipses~\cite{LQ2020}.
Curiously, in the absence of LFV signals the LQ expected best fit point (in orange) moves closer to the SM prediction, although overlapping with the $1\,\sigma$ contour of the extrapolated Belle II sensitivity.
On the other hand, should future measurements coincide with the central value of the current global average (with improved accuracy), a $V_1$-LQ explanation of the $\rdrds$ anomalies would be in conflict with future bounds on LFV processes (again in the absence of any LFV discovery at Belle II).
Thus, the evolution of future measurements of $\rdrds$ will prove instrumental in falsifying the vector LQ hypothesis. 

\section{Conclusion and outlook}
In recent years, numerous hints for the presence of LFU violation in charged and neutral current semi-leptonic $B$-meson decays have emerged in association with several observables.
Current EFT analyses appear to favour minimal models that can simultaneously address the tensions in both channels, due to a preference of {\it universal} contributions to $C_9^{bs\ell\ell}$ at the $b$-quark mass scale, which can be induced by RG running effects.
This interpretation is further strengthened by the very recently updated measurement of $\rk$.
Following the recent LHCb measurement~\cite{Aaij:2021vac}, we have updated fits of several NP hypotheses leading to a good agreement with the data.

In view of the absence of tree-level couplings between down-type quarks and neutrinos, 
$SU(2)_L$-singlet vector LQs are excellent candidates for a combined explanation of the $B$-meson decay anomalies, 
although subject to stringent constraints from LFV observables.
We have shown that a non-unitary flavour structure of the LQ couplings to SM matter is necessary in order to comply with numerous bounds from flavour observables, which have been measured to be in good agreement with the SM; such a structure can be generated via mixings of SM leptons with heavy vector-like doublet states.
We have further explored the flavour phenomenology of this simple vector LQ model , carrying a dedicated statistical analysis; this allowed identifying several ``{\it golden modes}'' that have excellent chances to be observed by upcoming experiments in the near future.
Finally, we highlighted the importance of future $\rdrds$ measurements for the model's viability.

\section*{Acknowledgements}
JK is indebted to the Organisers of the ``55$^\text{th}$ Rencontres de Moriond'' for the kind invitation, and grateful to his collaborators.
This project has received support from the European Union’s Horizon 2020 research and innovation programme under the Marie Skłodowska -Curie grant agreement No 860881-HIDDeN.

\appendix
\section{Additional updates on global fits}
\label{app:fits}
In this Appendix we provide several fit updates on additional NP hypotheses. These results were first presented in the Electroweak Session of the ``55$^\text{th}$ Rencontres de Moriond'', 24 March 2021~\cite{talk}.
Model-independent fits by other groups have since appeared in~\cite{fits}, while updated studies on $V_1$ leptoquark scenarios can be found in~\cite{LQ2021}.
All observables taken into account here can be found in Appendix~B of Ref.~\cite{LQ2020}.

As a first part of the present update, 
in Tables~\ref{tab:1d} and~\ref{tab:6d} we present our fit's results for hypotheses with NP contributions present in a single Wilson coefficient, as well as one hypothesis in which we consider 6 Wilson coefficients simultaneously. 
In order to establish a baseline, 
we first begin by giving our results prior to and after 
the inclusion of the updated LHCb measurement of $\rk$.
\begin{table}[h!]

    \centering
    \begin{tabular}{|c||c|c||c|c|}
    \hline
    & {pre Moriond 2021} & & post Moriond 2021 & \\
    \hline
    { WC} & BF $\pm 1\,\sigma$ & $\mathrm{Pull}_\mathrm{SM}$ & BF $\pm 1\,\sigma$ & $\mathrm{Pull}_\mathrm{SM}$\\
    \hline
    $\Delta C_9^{bs\mu\mu}$ & $-0.94 \pm 0.15$ & $5.9$ & $-0.89 \pm 0.14$ & ${6.3}$\\
    \hline
    $\Delta C_{10}^{bs\mu\mu}$ & $0.50 \pm 0.13$ & $4.0$ & $0.52 \pm 0.12$ & ${4.5}$\\
    \hline
    $\Delta C_9^{bs\mu\mu} = - \Delta C_{10}^{bs\mu\mu}$ & $-0.47 \pm 0.09$ & $5.5 $ & $-0.41^{+0.07}_{-0.08}$ & ${5.9}$\\
    \hline
    \end{tabular}
    \caption{Different $1$-d NP hypotheses leading to improvements with respect to the SM predictions. The pulls slightly increase upon inclusion of the new $\rk$ measurement.}
    \label{tab:1d}
\end{table}

\begin{table}[]
    
    \centering
    \hspace*{-3mm}\begin{tabular}{|r||c|c|c|c|c|c|c|}
    \hline
    & $C_9$ & $C_{10}$ & $C_9'$ & $C_{10}'$ & $C_7$ & $C_7'$ &$\mathrm{Pull}_\mathrm{SM}$ \\
    \hline
    pre Moriond 2021 & $-1.15^{+0.16}_{-0.15}$ & $0.17^{+0.15}_{-0.14}$ & $0.39^{+0.32}_{-0.33}$ & $-0.26^{+0.18}_{-0.17}$ & $0.001^{+0.014}_{-0.014}$ & $0.005^{+0.014}_{-0.014}$ & $6.2$\\
    \hline
    \hline
    post Moriond 2021 & $-1.15^{+0.16}_{-0.15}$ & $0.17^{+0.15}_{-0.14}$ & ${0.40}^{+0.32}_{-0.33}$ & $-0.26^{+0.18}_{-0.17}$ & $0.001^{+0.014}_{-0.014}$ & $0.005^{+0.014}_{-0.014}$ & ${6.5}$\\
    \hline
    \end{tabular}
    \caption{$6$-d NP hypothesis leading to an improvement with respect to the SM predictions. The best fit point mostly remains unaffected, while the significance slightly increases upon inclusion of the new data.}
    \label{tab:6d}
\end{table}

\begin{figure}[h!]
    \centering
    \includegraphics[width=0.5\textwidth]{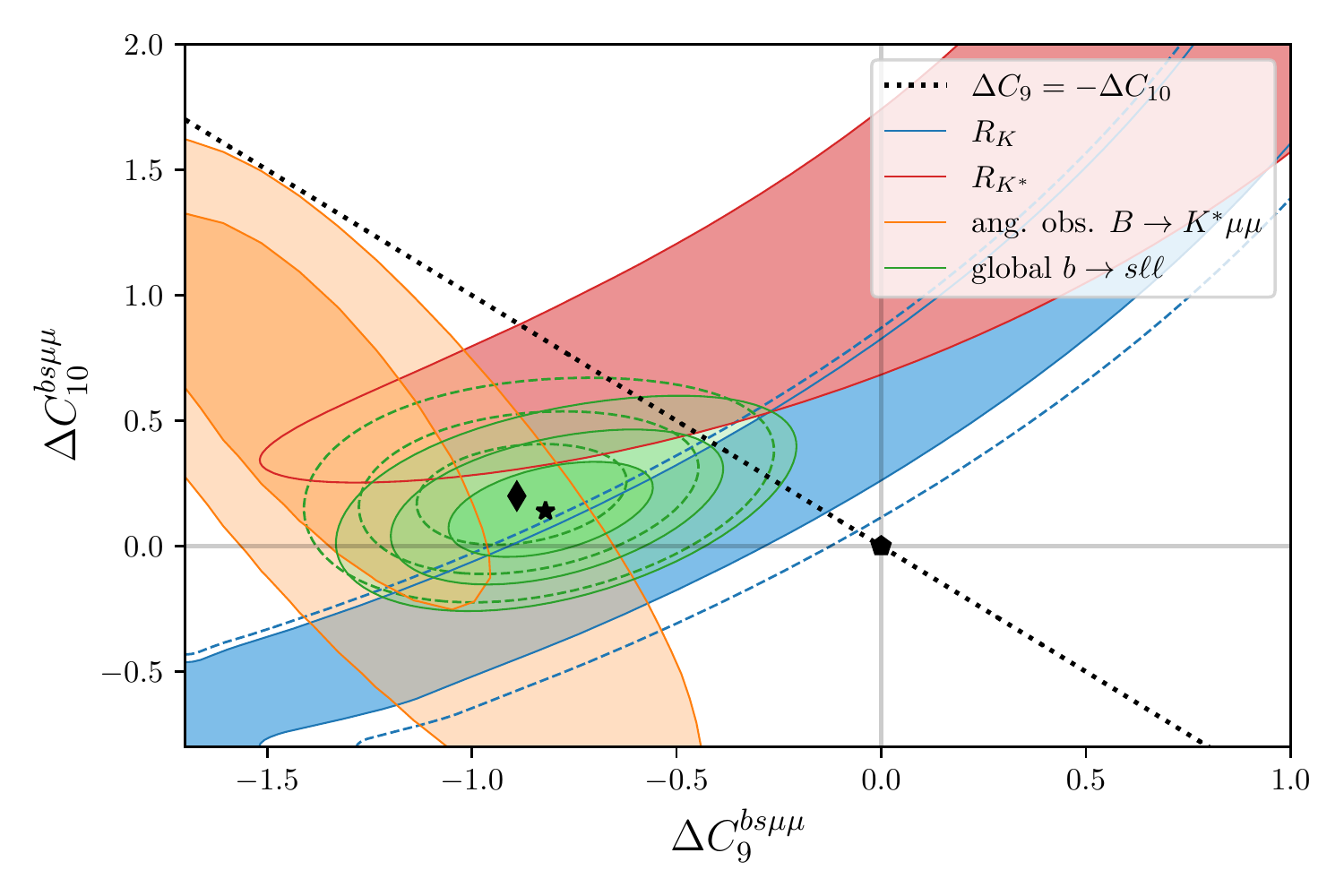}\includegraphics[width=0.5\textwidth]{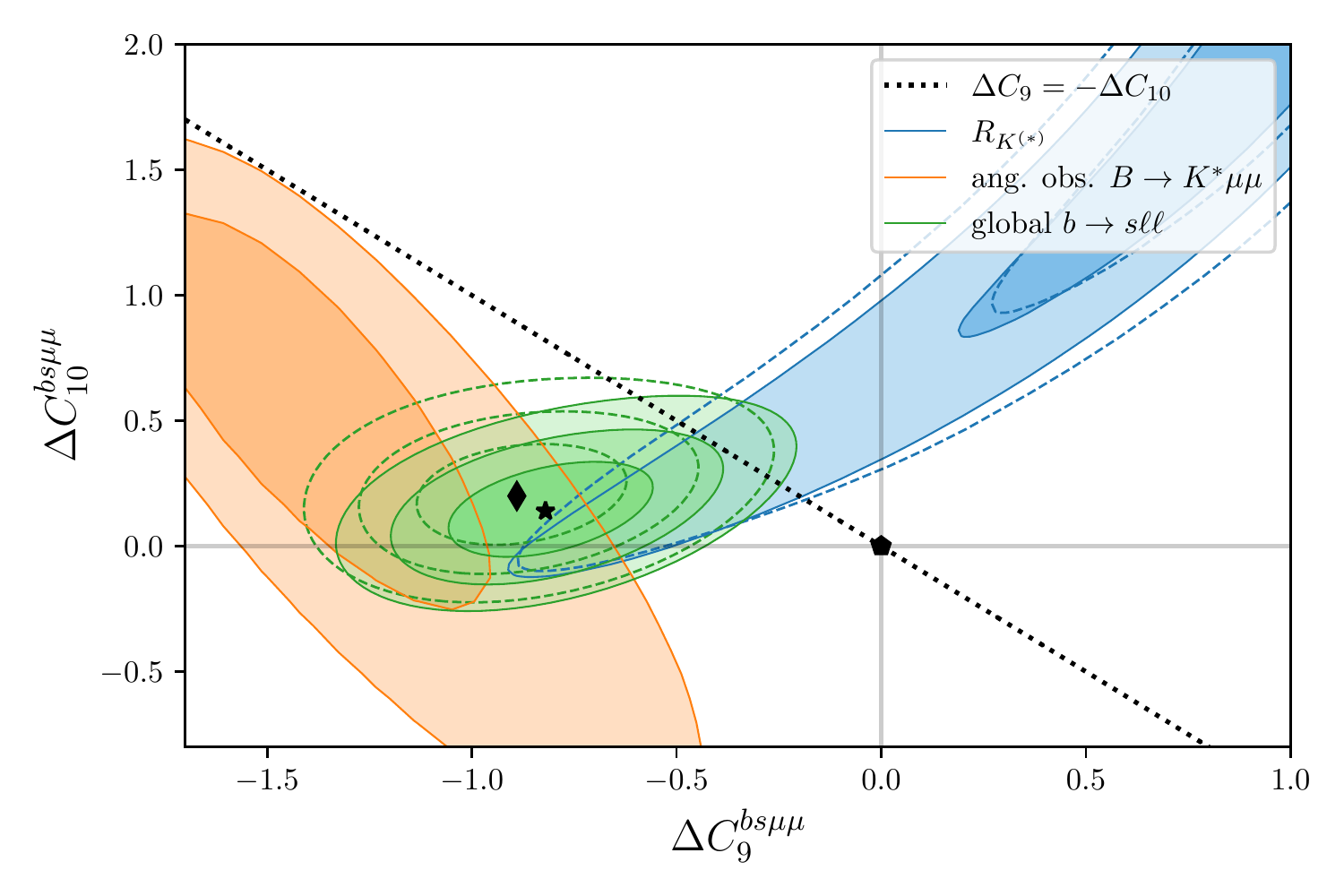}
    \caption{Likelihood contours in the plane of $\Delta C_9^{bs\mu\mu}$ vs. $\Delta C_{10}^{bs\mu\mu}$. The previous best fit point is given by $\Delta C_9^{bs\mu\mu} = -0.89^{+0.17}_{-0.16},\,\Delta C_{10}^{bs\mu\mu} = 0.20^{+0.13}_{-0.13}$, leading to a pull of $5.8\,\sigma$. Upon inclusion of the updated $\rk$ measurement, we obtain a new best fit point  $\Delta C_9^{bs\mu\mu} = {-0.82}^{+0.17}_{-0.16},\,\Delta C_{10}^{bs\mu\mu} = {0.14}^{+0.12}_{-0.12}$, leading to a slightly increased pull of $6.2\,\sigma$. The experimental measurement of $\rks < \rk$ leads to a slight tension that cannot be fully alleviated. Furthermore, there is a mild tension between $\rkrks$ and the data on the angular observables. Figures taken from Ref.~\cite{talk}.}
    \label{fig:C9_C10}
\end{figure}

\begin{figure}[h!]
    \centering
    \includegraphics[width=0.5\textwidth]{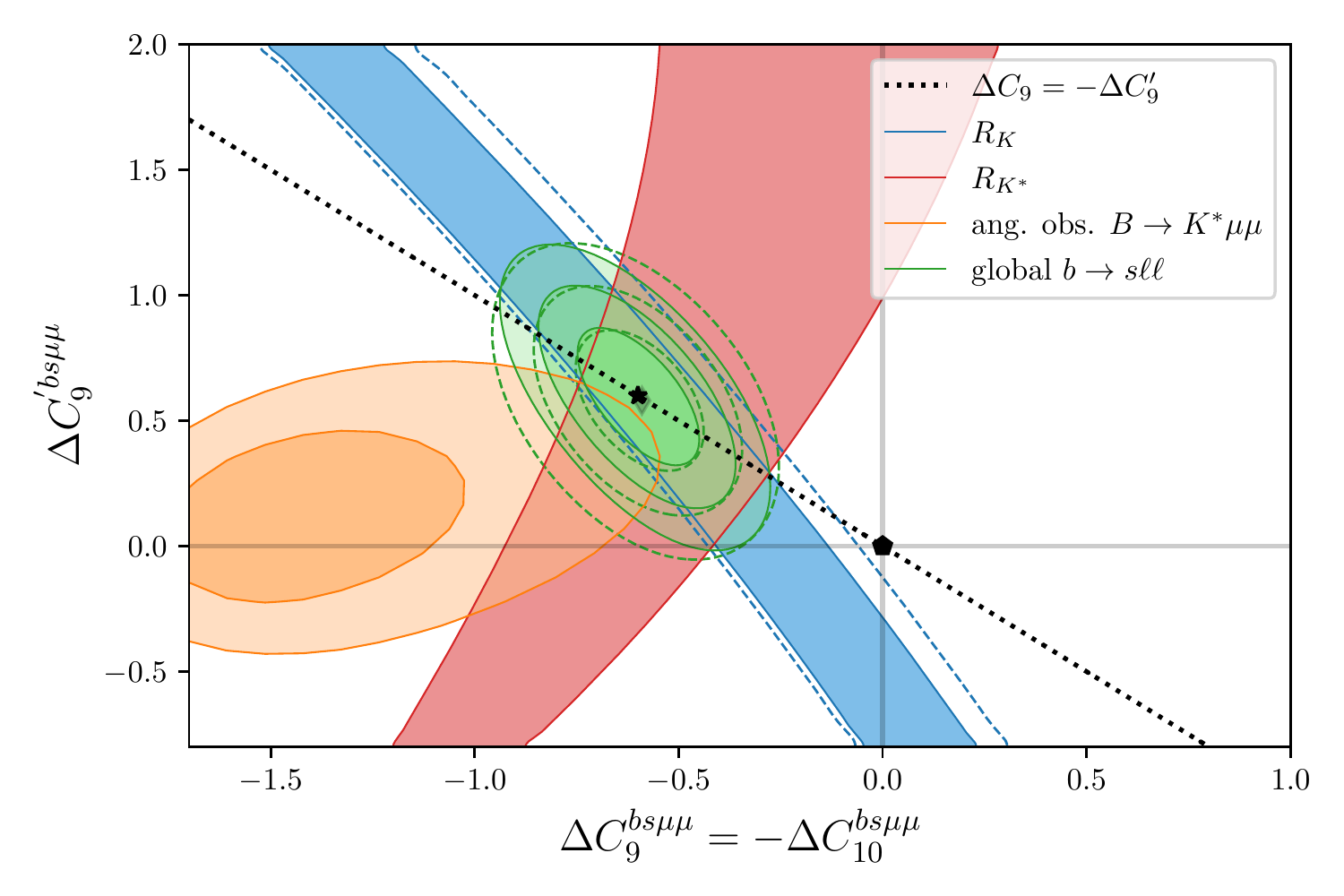}\includegraphics[width=0.5\textwidth]{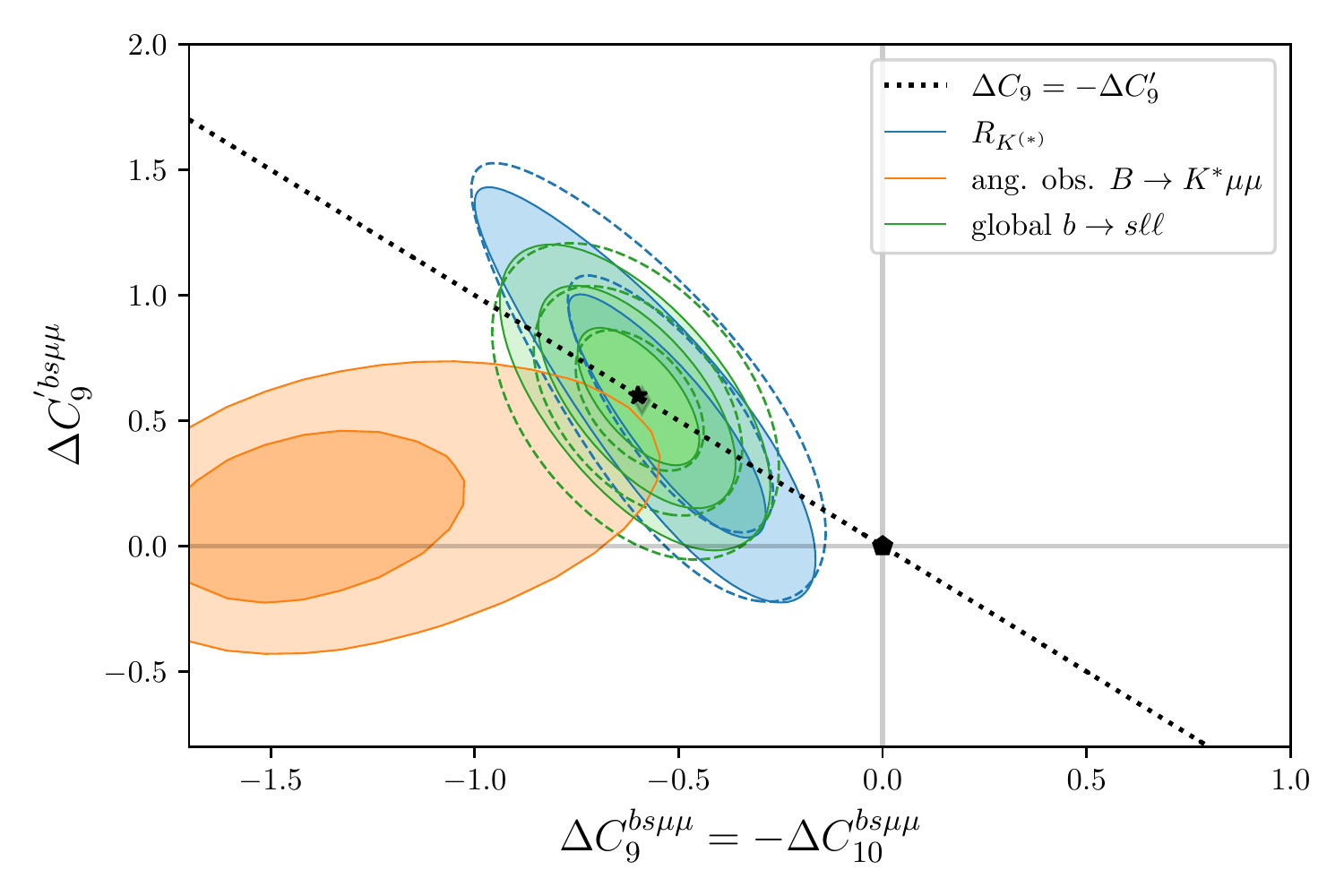}
    \caption{Likelihood contours in the plane of $\Delta C_9^{bs\mu\mu} = -\Delta C_{10}^{bs\mu\mu}$ vs. $\Delta C_{9}^{\prime bs\mu\mu}$. The previous best fit point is given by $\Delta C_9^{bs\mu\mu} = -0.59^{+0.10}_{-0.10},\,\Delta C_{9}^{\prime bs\mu\mu} = 0.58^{+0.18}_{-0.18}$, leading to a pull of $6.0\,\sigma$. Upon inclusion of the updated $\rk$ measurement, we obtain a new best fit point $\Delta C_9^{bs\mu\mu} =  -0.60^{+0.10}_{-0.10},\,\Delta C_{9}^{\prime bs\mu\mu} = 0.60^{+0.18}_{-0.18}$, leading to a slightly increased pull of $6.4\,\sigma$. The experimental measurement of $\rks < \rk$ can be fully accommodated if $\Delta C_9 \simeq C_9^\prime$. This however leads to a small tension between $\rkrks$ and the angular data. Figures taken from Ref.~\cite{talk}.}
    \label{fig:C910mu_C9pmu}
\end{figure}

\begin{figure}[h!]
    \centering
    \includegraphics[width=0.5\textwidth]{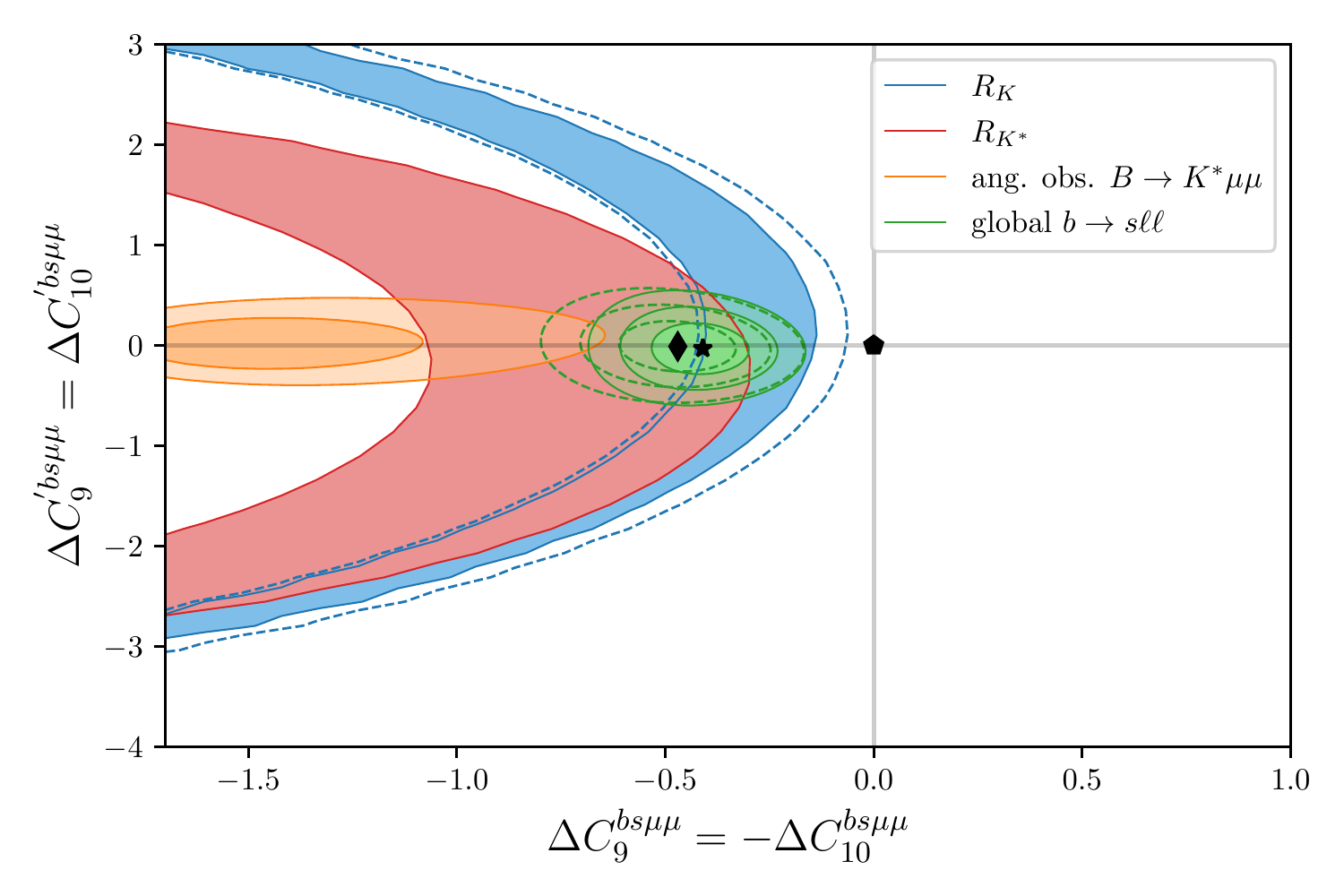}\includegraphics[width=0.5\textwidth]{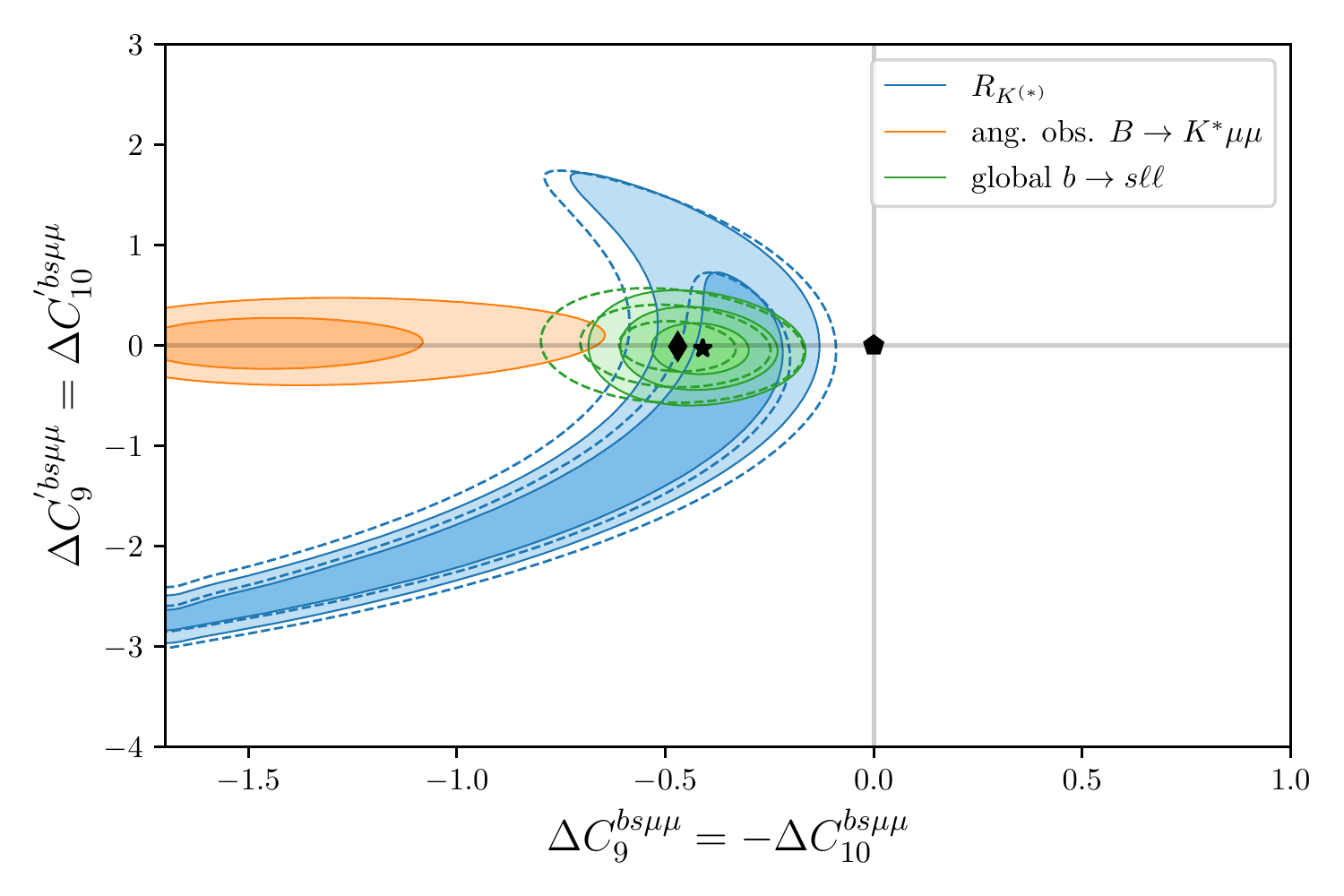}
    \caption{Likelihood contours in the plane of $\Delta C_9^{bs\mu\mu} = -\Delta C_{10}^{bs\mu\mu}$ vs. $\Delta C_{9}^{\prime bs\mu\mu} = \Delta C_{10}^{\prime bs\mu\mu}$. The previous best fit point is given by $\Delta C_9^{bs\mu\mu} = -0.47^{+0.09}_{-0.09},\,\Delta C_{9}^{\prime bs\mu\mu} = -0.01^{+0.17}_{-0.16}$, leading to a pull of $5.8\,\sigma$. Upon inclusion of the updated $\rk$ measurement, we obtain a new best fit point  $\Delta C_9^{bs\mu\mu} =  -0.41^{+0.07}_{-0.08},\,\Delta C_{9}^{\prime bs\mu\mu} = -0.03^{+0.17}_{-0.16}$, leading to a slightly {\it decreased} pull of $5.5\,\sigma$. The tensions between $\rk$ and $\rks$, and $\rk$ and the angular data, increases due to the improved accuracy in the experimental measurement of $\rk$, hence leading to a decreased pull. Figures taken from Ref.~\cite{talk}.}
    \label{fig:C910mu_C910pmu}
\end{figure}

In the remainder of this appendix we update the fits on NP hypotheses leading to new contributions to two (independent) Wilson coefficients. 
In contrast to the WET and SM-EFT hypotheses presented in Figure~\ref{fig:fits}, in all of the remaining scenarios 
$\rks < \rk$ be accommodated can only be accommodated for 
very specific configurations of the Wilson coefficients.
Thus, in Figures~\ref{fig:C9_C10},~\ref{fig:C910mu_C9pmu} and~\ref{fig:C910mu_C910pmu} we show in the left panels separate likelihood contours for $\rk$ and $\rks$, combining both observables in the right panels.
Descriptions of the results, including of the previous and current best fit points with profiled uncertainties, are given in the corresponding captions. 


\begin{thebibliography}{99}
{\small
%\cite{Zyla:2020zbs}
\bibitem{PDG}
P.~A.~Zyla \textit{et al.} [Particle Data Group],
%``Review of Particle Physics,''
PTEP \textbf{2020} (2020) no.8, 083C01.
% doi:10.1093/ptep/ptaa104

\bibitem{Belle:2019rba}
G.~Caria \textit{et al.} [Belle],
%``Measurement of $\mathcal{R}(D)$ and $\mathcal{R}(D^*)$ with a semileptonic tagging method,''
Phys. Rev. Lett. \textbf{124} (2020) no.16, 161803
% doi:10.1103/PhysRevLett.124.161803
[arXiv:1910.05864 [hep-ex]].

\bibitem{rkrks}
R.~Aaij \textit{et al.} [LHCb],
%``Search for lepton-universality violation in $B^+\to
%K^+\ell^+\ell^-$ decays,'' 
Phys. Rev. Lett. \textbf{122} (2019) no.19, 191801
%%doi:10.1103/PhysRevLett.122.191801
[arXiv:1903.09252 [hep-ex]];
R.~Aaij \textit{et al.} [LHCb],
%``Test of lepton universality with $B^{0} \rightarrow
%K^{*0}\ell^{+}\ell^{-}$ decays,'' 
JHEP \textbf{08} (2017), 055
%%doi:10.1007/JHEP08(2017)055
[arXiv:1705.05802 [hep-ex]].


\bibitem{angular}
R.~Aaij \textit{et al.} [LHCb],
%``Angular analysis of the $B^{0} \to K^{*0} \mu^{+} \mu^{-}$ decay
%using 3 fb$^{-1}$ of integrated luminosity,'' 
JHEP \textbf{02} (2016), 104
%%doi:10.1007/JHEP02(2016)104
[arXiv:1512.04442 [hep-ex];
R.~Aaij \textit{et al.} [LHCb],
%``Measurement of $CP$-Averaged Observables in the $B^{0}\rightarrow
%K^{*0}\mu^{+}\mu^{-}$ Decay,'' 
Phys. Rev. Lett. \textbf{125} (2020) no.1, 011802
%%doi:10.1103/PhysRevLett.125.011802
[arXiv:2003.04831 [hep-ex]];
R.~Aaij \textit{et al.} [LHCb],
%``Angular analysis of the $B^{+}\rightarrow K^{\ast+}\mu^{+}\mu^{-}$ decay,''
arXiv:2012.13241 [hep-ex].

%\cite{Aaij:2021vac}
\bibitem{Aaij:2021vac}
R.~Aaij \textit{et al.} [LHCb],
%``Test of lepton universality in beauty-quark decays,''
arXiv:2103.11769 [hep-ex].

\bibitem{LQ2020}
C.~Hati, J.~Kriewald, J.~Orloff, A.~M.~Teixeira, ``The fate of vector leptoquarks: impact of future flavour data,'' arXiv:2012.05883 [hep-ph].

\bibitem{talk}
J.~Kriewald,
Presentation at the ``$55^\text{th}$ Renconctres de Moriond'', 24 March 2021
\href{http://moriond.in2p3.fr/2021/EW/slides/3_flavour_03_kriewald.pdf}{http://moriond.in2p3.fr/2021/EW/slides/3\_flavour\_03\_kriewald.pdf}

\bibitem{Crivellin:2018yvo}
A.~Crivellin, C.~Greub, D.~M\"uller and F.~Saturnino,
%``Importance of Loop Effects in Explaining the Accumulated Evidence
%for New Physics in B Decays with a Vector LQ,'' 
Phys. Rev. Lett. \textbf{122} (2019) no.1, 011805
%%doi:10.1103/PhysRevLett.122.011805
[arXiv:1807.02068 [hep-ph]].

\bibitem{Hati:2019ufv}
C.~Hati, J.~Kriewald, J.~Orloff and A.~M.~Teixeira,
%``A nonunitary interpretation for a single vector LQ combined
%explanation to the $B$-decay anomalies,'' 
JHEP \textbf{12} (2019), 006
%%doi:10.1007/JHEP12(2019)006
[arXiv:1907.05511 [hep-ph]].

\bibitem{mutoe}
R.~Abramishvili \textit{et al.} [COMET],
%``COMET Phase-I Technical Design Report,''
PTEP \textbf{2020} (2020) no.3, 033C01
% doi:10.1093/ptep/ptz125
[arXiv:1812.09018 [physics.ins-det]];
%46 citations counted in INSPIRE as of 10 Dec 2020
L.~Bartoszek \textit{et al.} [Mu2e],
``Mu2e Technical Design Report,''
%%doi:10.2172/1172555
arXiv:1501.05241 [physics.ins-det].

\bibitem{Kou:2018nap}
E.~Kou \textit{et al.} [Belle II],
%``The Belle II Physics Book,''
PTEP \textbf{2019} (2019) no.12, 123C01
[erratum: PTEP \textbf{2020} (2020) no.2, 029201]
%%doi:10.1093/ptep/ptz106
[arXiv:1808.10567 [hep-ex]].

\bibitem{fits}
L.~S.~Geng, B.~Grinstein, S.~J\"ager, S.~Y.~Li, J.~Martin Camalich and R.~X.~Shi,
%``Implications of new evidence for lepton-universality violation in $b\to s\ell^+\ell^-$ decays,''
arXiv:2103.12738 [hep-ph];
W.~Altmannshofer and P.~Stangl,
%``New Physics in Rare B Decays after Moriond 2021,''
arXiv:2103.13370 [hep-ph].

\bibitem{LQ2021}
A.~Angelescu, D.~Be\v{c}irevi\'c, D.~A.~Faroughy, F.~Jaffredo and O.~Sumensari,
%``On the single leptoquark solutions to the $B$-physics anomalies,''
arXiv:2103.12504 [hep-ph];
G.~Hiller, D.~Loose and I.~Ni\v{s}and\v{z}i\'c,
%``Flavorful leptoquarks at the LHC and beyond: Spin 1,''
arXiv:2103.12724 [hep-ph];
C.~Cornella, D.~A.~Faroughy, J.~Fuentes-Mart\'\i{}n, G.~Isidori and M.~Neubert,
%``Reading the footprints of the B-meson flavor anomalies,''
arXiv:2103.16558 [hep-ph].
}

\end{thebibliography}
\end{document}